# Studies of Multiferroic System of $LiCu_2O_2$ II
## Magnetic Structures of Two Ordered Phases with Incommensurate Modulations

Yoshiaki Kobayashi, Kenji Sato, Yukio Yasui, Taketo Moyoshi, Masatoshi Sato*, and Kazuhisa Kakurai[1]

*Department of Physics, Division of Material Science, Nagoya University, Furo-cho, Chikusa-ku Nagoya 464-8602*

*[1]Quantum Beam Science Directorate, Japan Atomic Energy, Tokai, Ibaraki 319-1195*

Neutron diffraction and $^7$Li-NMR have been applied to determine the multiferroic system $LiCu_2O_2$, which has four chains (ribbon chains) of edge-sharing $CuO_4$ square planes in a unit cell. We have confirmed that there are successive magnetic transitions at $T_{N1}$=24.5 K and $T_{N2}$=22.8 K. In the $T$ region between $T_{N1}$ and $T_{N2}$, the quasi one-dimensional spins ($S$=1/2) of $Cu^{2+}$ ions within a chain have a collinear and sinusoidally modulated structure with Cu-moments parallel to the $c$-axis and with the modulation vector along the $b$-axis. At $T < T_{N2}$, an ellipsoidal helical spin structure with the incommensurate modulation has been found. Here, we present detailed parameters, describing the modulation amplitudes, helical axis vectors as well as the relative phases of the modulations of four ribbon chains, which can well reproduce both the NMR and neutron results in the two magnetically ordered phases. This finding of the rather precise magnetic structures enables us to discuss the relationship between the magnetic structure and the multiferroic nature of the present system in zero magnetic field, as presented in our companion paper (paper I), and open a way how to understand the reported electric polarization under the finite magnetic field.

## 1. Introduction

Multiferroic systems that exhibit ferroelectric and magnetic simultaneous transition have been actively studied and various kinds of multiferroic system have been found.[1)] Among these multiferroics, systems with large ferroelectric polarization attracts much interest from the view point of technical application, because their (electric polarizations)/(magnetic ordering patterns) may be controlled by the external magnetic/electric fields. Apart from such systems, we have been searching spin $S$ = 1/2 multiferroic systems, expecting possible quantum effects on the multiferroic nature, and found a new multiferroic system of $LiVCuO_4$,[2, 3)] which has quasi one-dimensional $CuO_2$ chains ($CuO_2$ ribbon chains) formed of the edge-sharing $CuO_4$ square plaquettes. Its magnetic structure is helical with the modulation vector $\boldsymbol{Q}$ along the chain direction, and as theoretically predicted,[4, 5)] it exhibits a ferroelectric transition with the spontaneous polarization $\boldsymbol{P}$ described by the relation $\boldsymbol{P} \propto \boldsymbol{Q} \times \boldsymbol{e}_3$, where $\boldsymbol{e}_3$ is the helical axis. In addition to this system, we have also found that $Na_2Cu_3(GeO_3)_4$, which has $Cu_3O_8$ clusters formed of three $CuO_2$ square planes, is also a multiferroic.[6)]

$LiCu_2O_2$ is another example of multiferroic systems with $CuO_2$ ribbon chains reported by Park *et al.*[7)] The crystal structure is orthorhombic (space group *Pnma*), as shown in Fig. 1 of the companion paper (paper I),[8)] and there are four ribbon chains within a unit cell.[9)] The Cu spins have the ordering with the incommensurate modulation[10)] accompanied by the ferroelectricity.[7)] However, to understand its ferroelectricity, there are various issues which have to be understood, as stated in paper I,[8)] mainly because the magnetic structure is ambiguous, in spite of many trial works to determine the magnetic structure.[10-12)]

Here, systematic studies have been carried out, using samples prepared and characterized by ourselves,[8)] to understand the multiferroic behaviors of this system. In the course of the studies, we have first realized that for the determination of its magnetic structures, it is insufficient to reproduce only one kind of experimental data by choosing a set of parameters, because we have to use a number of parameters to describe the modulated magnetic structures of the four ribbon chains within a unit cell. Actually, a magnetic structure proposed by NMR measurements[11)] is different from the one proposed by neutron scattering studies.[10)] We have therefore applied various methods including $^7$Li-NMR and neutron magnetic scattering on single crystals, trying to consistently explain data obtained by these methods. We have first confirmed that the system has two successive magnetic transitions at $T_{N1} \cong 24.5$ K and $T_{N2} \cong 22.8$ K.[12)] In the temperature region between $T_{N1}$ and $T_{N2}$, the magnetic structure has been found to be a sinusoidally modulated incommensurate one. Below $T_{N2}$, a helical ellipsoidal structure with the incommensurate modulation has been obtained. The present paper reports these magnetic structure determinations of $LiCu_2O_2$ carried out mainly by $^7$Li-NMR and neutron magnetic scattering measurements on single-crystal samples.

* corresponding author: e43247a@nucc.cc.nagoya-u.ac.jp

## 2. Experiments

We have carried out $^7$Li-NMR and neutron scattering measurements using single-crystal samples, for which details of their characterization and macroscopic properties are presented in paper I.[8)] The $^7$Li-NMR measurements were carried out by the standard coherent pulse method: The $^7$Li-NMR spectra were measured by recording the integrated intensity $I_n$ of the nuclear spin-echo signal at the NMR frequency of $f$ = 33.14 MHz with the applied magnetic field $H$ being changed stepwise. The $T_2$-corrction was made by measuring the $\tau$-dependence of $I_n$ at various fixed values of $H$ to obtain the value of $I_n$ ($\tau\rightarrow 0$) where $\tau$ is the interval time between the first and second RF-pulses. In these measurements, minute measurements were required to see the significantly modulated structure of $I_n$ as a function of $\tau$ produced by the interference of the three $eqQ$-split lines at $\omega_0 \equiv \gamma(H + H_{int})$ and $\omega_0 \pm\nu_Q$ ($\nu_Q$: nuclear quadrupole frequency).

Neutron scattering measurements were carried out on a single crystal of $^7LiCu_2O_2$ using the triple axis spectrometer TAS-1 installed at JRR-3 of JAEA in Tokai. To avoid the large neutron absorption of Li nuclei, we used the $^7$Li isotope in the measurements. The horizontal collimations were 40'-40'-80'-open, and the neutron wave length was 2.359 Å. The 002 reflection of Pyrolytic graphite (PG) was used as the monochromator. A PG filter was placed after the second collimator to suppress the higher-order contamination. The crystal was oriented with the [100] and [010] axes, in one case, and the [102] and [010] ones, in another case, in the scattering plane. The size of the crystals was ~18×16×1mm$^3$, and the integrate intensities of 20 nonequivalent magnetic Bragg reflections were measured. The crystal was set in an Al-can filled with exchange gas, and the can was attached to the cold head of the Displex type refrigerator. In the analyses of the data, the anisotropic magnetic form factor for $Cu^{2+}$ ions deduced from the $x^2$-$y^2$ orbit[13)] was used, and the absorption corrections considering the shape of the single crystal were made.

## 3. Experimental Results and discussion

### *3. 1 Confirmation of the existence of successive transitions*

In the neutron diffraction measurements on a single crystal of $LiCu_2O_2$ at low temperatures, we found Bragg reflections at several $\boldsymbol{Q}$ points of ($h$/2, $k\pm\delta$, $l$) with $h$, $l$=odd and $k$=integer in the reciprocal lattice units of $LiCu_2O_2$. These reflections appear, as shown in Fig. 1(a) for the 0.5 δ 1 reflection, at the temperature $T_{N1}$, which coincides with the temperatures of the anomalies in the dielectric susceptibility ε and specific heat shown in our companion paper (paper I).[8)] We can consider these reflections to be magnetic, because the strong reflections can be found only in the small $Q$ region The δ value of 0.172-0.174 indicates that the magnetic ordering is characterized by the incommensurate modulation vectors, corresponding to the periodicity of (5.76-5.80)$b$ along the Cu-chain direction (see Fig. 1(b)). The period of the spin modulation along $a$-axis is 2$a$, and spins at $Cu^{2+}$ sites shifted by (1, 0, 0) are anti-parallel to those at the original sites.

As shown in Fig. 1(a), a weak but clear anomaly can be seen in the integrated intensity ($I$) -$T$ curve at the second transition temperature $T_{N2}$, where anomalies have been observed in the dielectric susceptibility ε-$T$ and specific heat ($C_s$)-$T$ curves, too.[8)] The existence of the magnetic structure change at $T_{N2}$ is also supported by the significant difference of the $^7$Li-NMR spectra between temperatures above and below $T_{N2}$, as can be found in Fig. 2, indicating that we have to determine the magnetic structures in both the intermediate ($T_{N2} < T < T_{N1}$) and low temperature ($T < T_{N2}$) phases.

### *3.2 Magnetic structure in the intermediate phase*

The integrated neutron Bragg intensities $I$ measured in the intermediate phase are plotted in Fig. 3(a) for several reflections against the values of the corresponding reflections calculated by the model fitting. In the model fitting, we have assumed that the magnetic moments are collinear and sinusoidally modulated, as described by the relation $m^z{}_i(y) = \mu_c\cdot\cos(\boldsymbol{Q}\cdot\boldsymbol{y} + \phi_i)$, $m^x{}_i = 0$, and $m^y{}_i = 0$, considering the results of X-ray magnetic scattering measurements,14) where $\boldsymbol{y}$ represents the positions of Cu2+ ions, the superscripts indicate the components of the ordered moments, and $\phi_i$ describes the relative phases of the $i$-th ($i$ =1~4) $CuO_2$ ribbon chains labeled Cu1-Cu4 in Fig. 3(b). Then, we have found, for the value of $\boldsymbol{Q}\cdot\boldsymbol{b}$ = $\Delta\phi$ ~62.03° known from the δ value, that a parameter set $\phi_1$ = 0°, ϕ2 = 90°, $\phi_3$ = 90° and $\phi_4$ = 180° can reproduce the observed results quite well with $\mu_c$=0.3 ± 0.1 $\mu_B$, as shown in Fig. 3(a). Another set $\phi_1$ = 0°, $\phi_2$ = –90°, $\phi_3$ = –90° and $\phi_4$ = –180° can also reproduce the results well with the same value of $\mu_c$. The latter set can be obtained by reversing the $b$-axis.

Then, in order to examine if the magnetic structure is correctly determined by the above analyses of the neutron data, we have studied the $^7$Li-NMR spectra, too. Figures 2(a)-2(c') show the NMR data taken at various phases. The left and right panels show the results for the external magnetic fields $\boldsymbol{H}$ || *a*- (or *b*-) axis and $\boldsymbol{H}$ || *c*-axis, respectively. (Note that the *a*- and *b*-axes cannot be distinguished due to the existence of the domains). At 25-26 K (> $T_{N1}$), Li atoms are equivalent and the observed spectra are just composed of three peaks split by the nuclear quadrupole interaction (*eqQ*) as expected for nuclei with spin $I$ = 3/2. The profile indicates that the nuclear electric quadrupole resonance frequencies $f_Q$ is very small, and the existence of these three peaks is the origin of the τ-modulation stated in section 2. (Because the period of the τ-modulation below $T_{N1}$ is almost equal to that above $T_{N1}$ and does not change with varying $T$ through $T_{N2}$, the change of the profile with $T$ found in Fig. 2 is not due to the change of the nuclear quadrupole interactions.)

When $T$ is lowered to the region $T_{N2} < T < T_{N1}$, the spectra broaden due to the internal fields induced at Li sites by the ordered moments of $Cu^{2+}$, and the profiles observed for both direction of $\boldsymbol{H}$ change to those characteristic of rather simple incommensurate modulations of the internal fields. To fit to these profiles for two $\boldsymbol{H}$ directions, we have used the parameters of the *c*-sinusoidal magnetic structure obtained above by the neutron diffraction studies, and calculated the distribution of the dipolar field induced at the Li sites by the Cu moments $\boldsymbol{m}_i=(m^x_i, m^y_i, m^z_i)$ using the equation.

$$\boldsymbol{H}_{\rm dip} = \Sigma_i(-\boldsymbol{m}_i/r^3 + 3(\boldsymbol{m}_i\cdot\boldsymbol{r})\boldsymbol{r}/r^5),$$

where $\boldsymbol{r}$ describes the position from the Li sites. In the actual calculations, the summation of $\boldsymbol{H}_{\rm dip}$ over the Cu moments within the neighboring 7 × 7 × 7 unit cells (For more than 3 × 3 × 3 unit cells, Hdip hardly changes.) are taken at each Li site, where we have found that the results can explain the profile for $\boldsymbol{H}$ || $\boldsymbol{a}$ (or $\boldsymbol{b}$) as shown by the gray line in Fig. 2(b)L. (In the calculation of the gray line, each spectrum was treated to have the Gaussian broadening of 0.006 T.) However, the results for $\boldsymbol{H}$ || $\boldsymbol{c}$ give much smaller width than the observed value. This discrepancy for $\boldsymbol{H}$ || $\boldsymbol{c}$ can be solved by considering the transferred hyperfine field $\boldsymbol{H}_{\rm hf}$ at each Li-site from two nearest neighboring Cu-moments. (Although two Cu-moments have the opposite directions, $\boldsymbol{H}_{\rm hf}$ is not negligible, because the distances from the two Cu-sites to the Li-site are slightly different.) We have found that to reproduce the absolute values of the distribution widths of the spectra, the difference ($A_{\rm hf}$–$A'_{\rm hf}$) between the hyperfine coupling constants $A_{\rm hf}$ and $A'_{\rm hf}$ of Li-nuclei and two nearest neighboring Cu-moments, is ~1 kOe/$\mu_B$. (This is ~1/10 of $A_{\rm hf}$ in the paramagnetic state estimated roughly from the $T$-dependence of Knight shift in $T > T_{N1}$.) For this value of $A_{\rm hf}$–$A'_{\rm hf}$, the calculated line shown by the gray line in Fig. 2(b)R can be obtained. The gray line of the left panel of Fig. 2(b)L is not changed by the consideration of the transferred hyperfine field, as far as the Cu moments are parallel and antiparallel to the *c*-axis. If we consider the direction of the modulated Cu moments to be within the *ab*-plane, we cannot explain the observed profiles, because in such cases, profiles are expected to have quite different characteristic from the observed ones. As the results of the present studies, the collinear ordering with the moment direction along $\boldsymbol{c}$ and with the incommensurate modulation along $\boldsymbol{b}$, have been found to consistently explain both the NMR and neutron data of the intermediate phase, where the amplitudes and relative phases of the modulations of the four chains are also given (see Fig. 3(b)).

### *3.3 Magnetic structure in the low T phase*

The NMR profiles in the low $T$ phase (< $T_{N2}$) are more complicated than that in the intermediate phase. Therefore, in order to find a simple way to determine the parameters to properly describe the magnetic structure, we first consider that the integrated intensities of the neutron magnetic reflections at 12 K (<$T_{N2}$) plotted against those of the corresponding reflections at 23.3 K ($T_{N2} < T < T_{N1}$) in the log-log scales do not exhibit significant deviations from the straight line. It indicates that the relative phases $\phi_i$ ($i$ =1~4) do not meaningfully change through $T_{N2}$. This idea is rationalized by the fact that if the relative phases $\phi_i$ are changed, the significant deviation of the above plot from the straight line is observed. Then, we have tried to use, in the calculations of the NMR profiles, the same values of $\phi_i$ ($i$ =1~4) determined in the intermediate phase, and adopted the helical structure described by the following relations,

$$m^x_i(y) = \mu_{ab}\cdot\sin(\boldsymbol{Q}\cdot\boldsymbol{y} + \phi_i)\cdot\sin\alpha,$$

$$m^y_i(y) = \mu_{ab}\cdot\sin(\boldsymbol{Q}\cdot\boldsymbol{y} + \phi_i)\cdot\cos\alpha,$$

$$m^z_i(y) = \mu_c\cdot\cos(\boldsymbol{Q}\cdot\boldsymbol{y} + \phi_i).$$

Then, considering both the dipolar and transferred hyperfine fields, we have obtained rather satisfactory fits to the observed profiles, as shown in Figs. 2(*c*)L and 2(c)R by the gray lines for both $\boldsymbol{H}$ || $\boldsymbol{a}$- (or $\boldsymbol{b}$-) axis and $\boldsymbol{H}$ || $\boldsymbol{c}$-axis, respectively. The parameters are summarized as follows. $\mu_{ab}$= 0.45±0.10 $\mu_B$, $\mu_c$= 0.85±0.15 $\mu_B$, $\alpha$ = –45° (and +135°), $\phi_1$ = 0, $\phi_2$ = 90°, $\phi_3$ = 90°, $\phi_4$ = 180° and ($A_{\rm hf}$–$A'_{\rm hf}$) ~1 kOe/$\mu_B$, where the two sets with different $\alpha$ values correspond to the opposite signs of the spin-rotation axis $\boldsymbol{e}_3$. Other two sets $\mu_{ab}$ = 0.45±0.10 $\mu_B$, $\mu_c$=0.85±0.15 $\mu_B$, $\alpha$ = +45° (and –135°), $\phi_1$ = 0, $\phi_2$ = –90°, $\phi_3$ = –90°, $\phi_4$ = –180° and ($A_{\rm hf}$–$A'_{\rm hf}$) ~1 kOe/$\mu_B$ are also possible. The magnetic patterns described by these parameters can be obtained by reversing the *b*-axis. To get the better fittings than the above results, we have introduced a 3***Q***-component with the modulation amplitude smaller that of the ***Q***-component by a factor of 1/6, and the results are shown in Figs. 2(c')L and 2(c')R, where the fittings shown by the gray lines seems to be improved.

To examine if the parameters determined above can reproduce the integrated intensities of neutron magnetic reflections, we have calculated the intensities for these parameters, and obtained the results shown in Fig. 4(b), where the result is found to be satisfactory. The magnetic reflection expected at the points corresponding to the 3***Q*** component is not large enough to be observed in the present experiments.

These magnetic structures have following characteristics. (1) The helical plane has a shape of an ellipsoid with the longer axis of ~0.85 $\mu_B$ and the shorter axis of ~0.45 $\mu_B$. (2) The helical axes are along [1 1 0], [-1 -1 0], [1 -1 0] or [-1 1 0] depending on the sets of the parameters. One of the obtained helical structures determined above is shown, for example, in Figs. 5(a) and 5(b) for $\alpha = -45°$, $\phi_1 = 0$, $\phi_2 = 90°$, $\phi_3 = 90°$, $\phi_4 = 180°$.

We have shown the magnetic structures in the low-*T* phase. There are four similar structures with equal or very close energies. Their coexistence may induce the domain structure of the system. Judging from the relation $\boldsymbol{P} \propto \boldsymbol{Q} \times \boldsymbol{e}_3$, the magnetic structures determined in this work are consistent with the absence of the spontaneous electric polarization in the intermediate phase, and the presence of the polarization along ***c*** in the low-*T* phase present firm bases of the arguments on the relationship between the multiferroic nature and the noncollinear modulated magnetic structures.

## 4. Conclusion

The magnetic structures of $LiCu_2O_2$ have been determined by the combined studies of $^7$Li-NMR and neutron scattering measurements. First, the existence of the successive magnetic transitions at 24.7 K (= $T_{N1}$) and 22.8 K (= $T_{N2}$) have been confirmed. In the intermediate phase ($T_{N2} < T < T_{N1}$), it has been found that the structure is collinear and sinusoidally modulated with the spins along ***c***. We have determined the parameters describing the modulated structure including the relative phases of the modulations of the four ribbon chains within a unit cell. In the low-*T* phase ($T < T_{N2}$), we have found that it has the incommensurate ellipsoidal helical structure with the helical axis tilted by ~45° from the *b*-axis within the *ab* plane. The longer axis of the ellipsoid is ~0.85 $\mu_B$ and the shorter is ~0.45 $\mu_B$. The relative phases of the modulations of the four ribbon chains within a unit cell have also been determined. The determination of the magnetic structures presents bases of arguments on the multiferroic nature of this quasi one dimensional quantum spins.

Acknowledgements

This work is supported by Grants-in-Aid for Scientific Research from the Japan Society for the Promotion of Science (JSPS) and Grants-in Aid on priority areas from the Ministry of Education, Culture, Sports, Science and Technology.

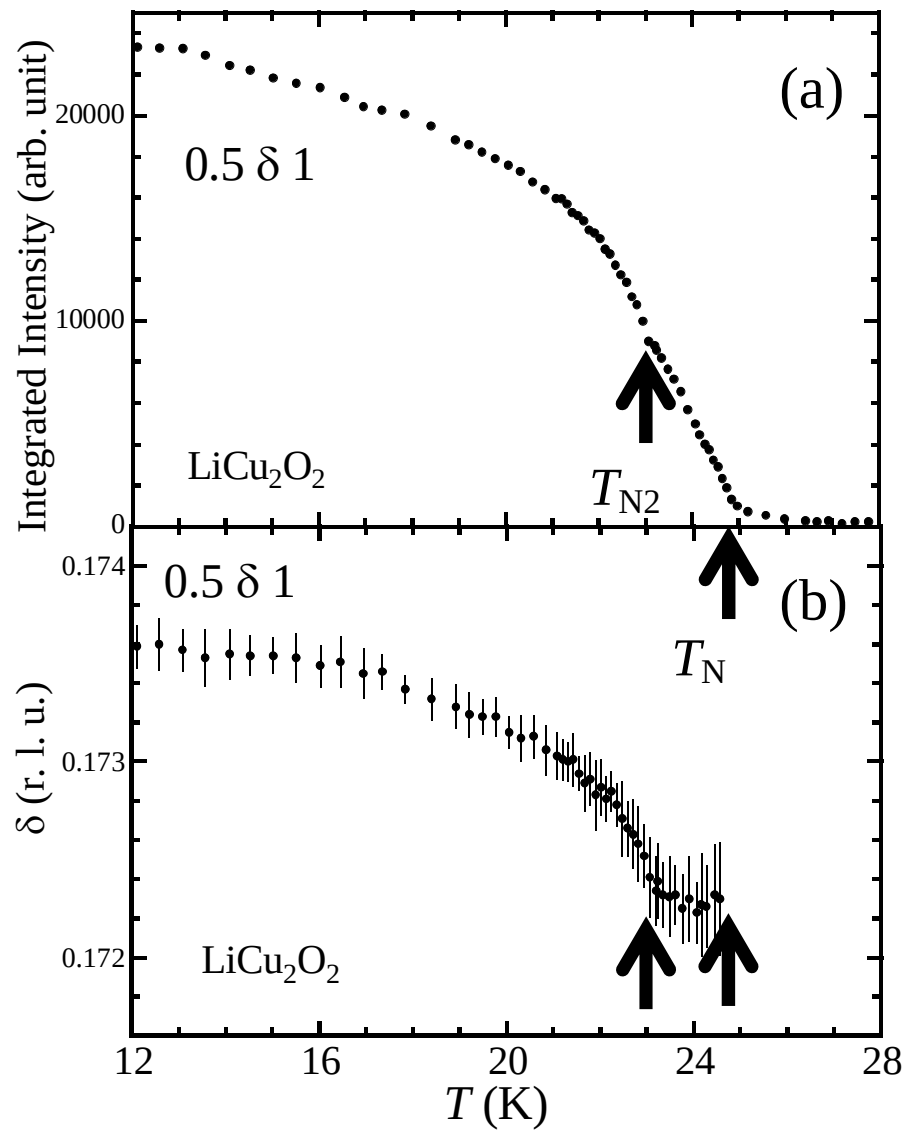

Fig. 1. Integrated intensity and the δ value of 0.5 δ 1 magnetic reflection observed for $LiCu_2O_2$ are shown against *T*.

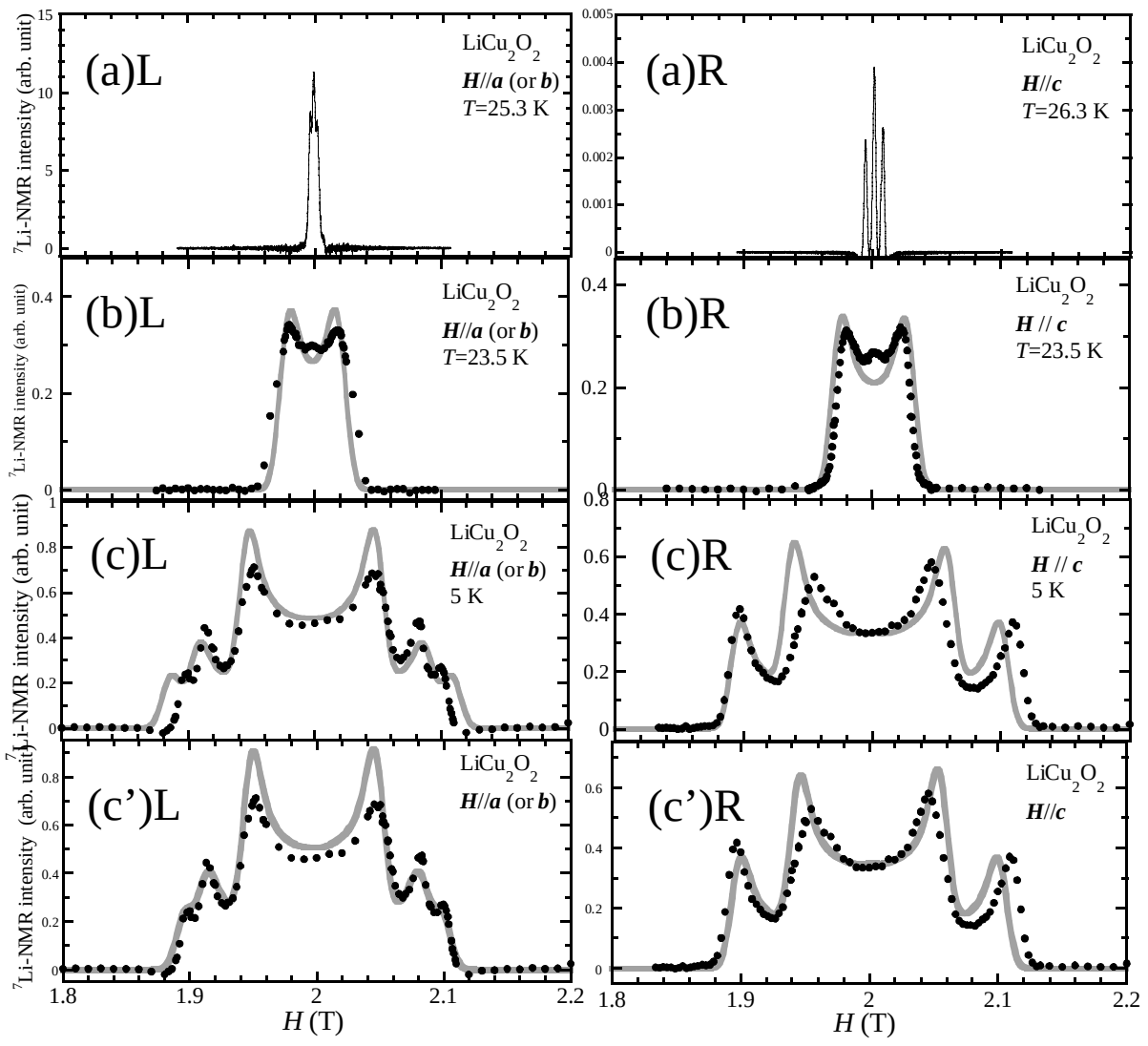

Fig. 2. [7]Li-NMR spectra of $LiCu_2O_2$ for the applied magnetic field $\boldsymbol{H}$ // $\boldsymbol{a}$ (or $\boldsymbol{b}$) (left side) and // $\boldsymbol{c}$ (right side) are shown in (a) (b) and (c and c') for the paramagnetic phase ($T > T_{N1}$), in the intermediate phase ($T_{N2} < T < T_{N1}$), and in the low-$T$ phase ($T < T_{N2}$), respectively. In (b) and (c and c'), the fitted results are also shown by the gray lines. In (c'), the fitted line is obtained by considering the small higher harmonic component of the spin modulation.

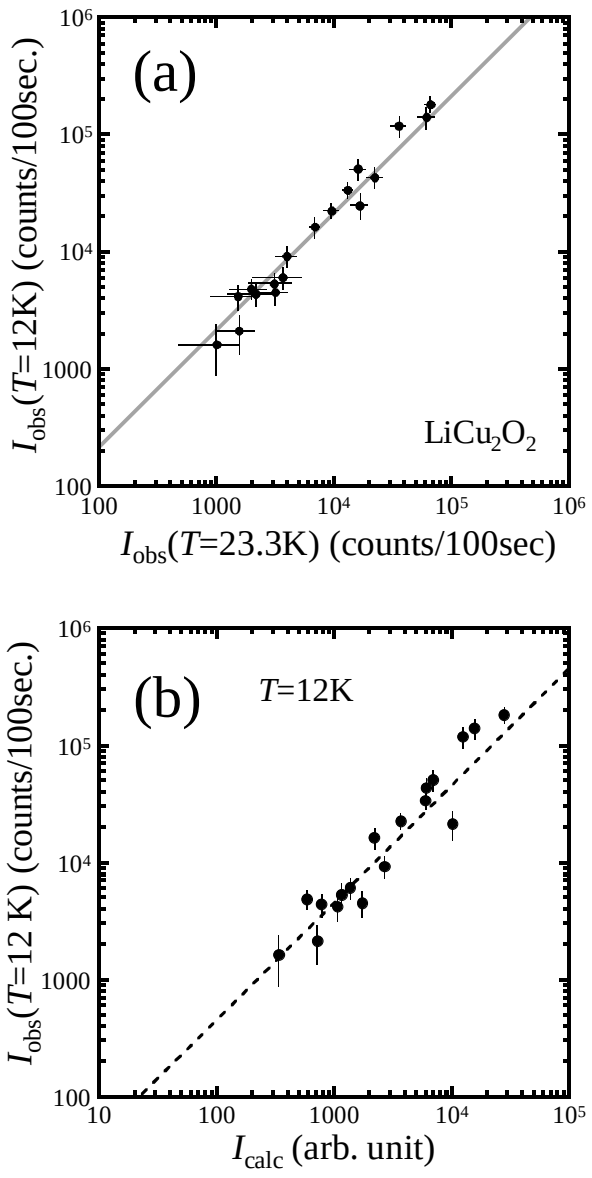

Fig. 3. (a) Results of the fitting to the observed integrated intensities of various magnetic reflections in the intermediate phase by a model of collinear and sinusoidally modulated structure with the Cu-moments parallel to $c$-axis, as shown in (b).

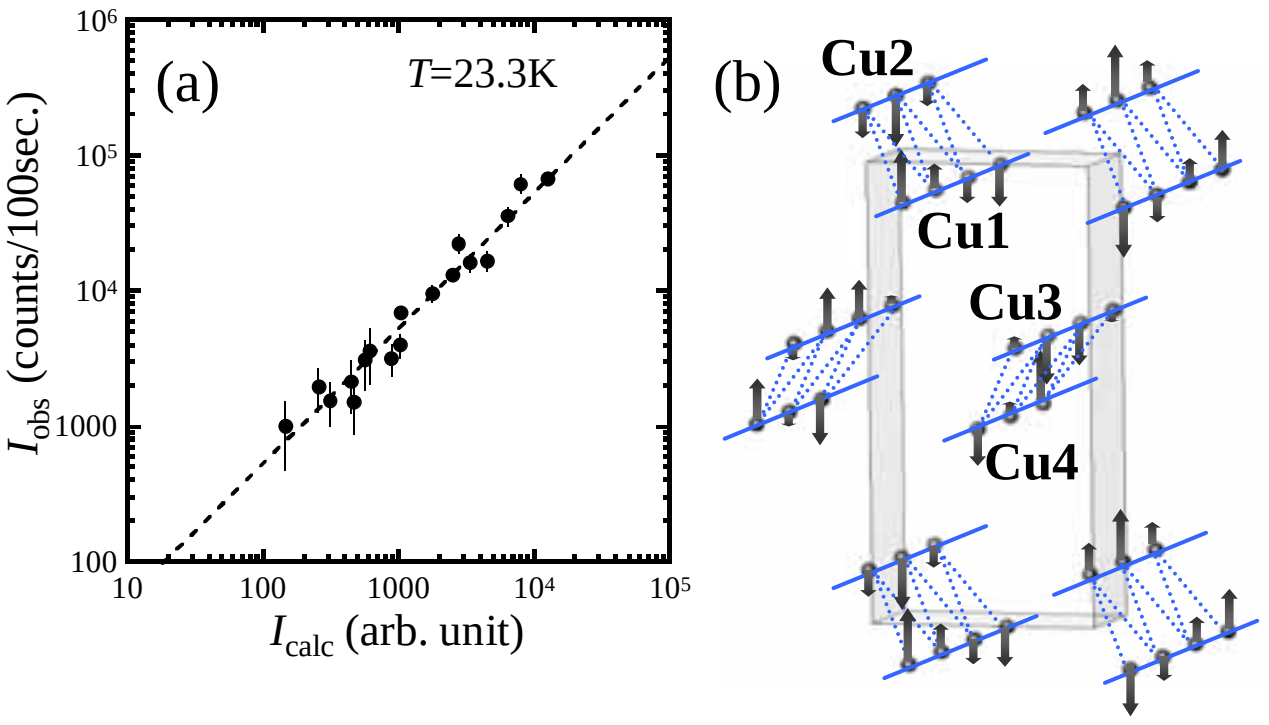

Fig. 4. (a) The observed integrated intensities of various magnetic reflections at 12 K are plotted against those of the corresponding reflections at 23.3 K. (b) The observed integrated intensities of the various magnetic reflections in the low-$T$ phase are plotted against those calculated using the parameters determined in the analyses of the NMR data shown in Fig. 2(c).

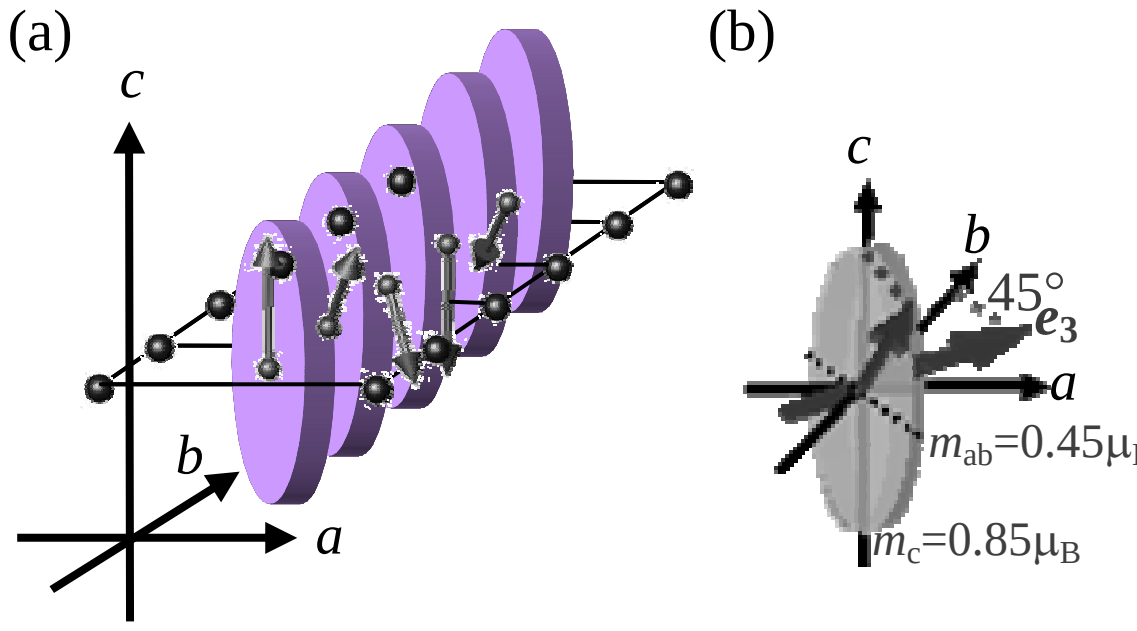

Fig. 5. (a) Magnetic structure on a $CuO_2$ ribbon chains in the low-$T$ phase. (b) The relationship between helical axis $\boldsymbol{e}_3$ (// 110) and modulation vector (// $\boldsymbol{b}$) is shown for one of possible parameter sets, $\mu_{ab}$= 0.45±0.10 $\mu_B$, $\mu_c$= 0.85±0.15 $\mu_B$, α= −45° (and +135°), $\phi_1$= 0, $\phi_2$= 90°, $\phi_3$= 90°, $\phi_4$= 180° and ($A_{hf}$–$A'_{hf}$) ~1 kOe/$\mu_B$.